\pdfoutput=1
\documentclass[runningheads]{llncs}
\usepackage[T1]{fontenc}
\usepackage{graphicx}
\usepackage{booktabs}
\usepackage{threeparttable}
\usepackage{multirow}
\usepackage{algorithmicx,algorithm}
\usepackage{amsbsy}
\usepackage{amsfonts}
\usepackage{marvosym}
\begin{document}
%
\title{DeU-Net: Deformable U-Net for 3D Cardiac MRI Video Segmentation}
\titlerunning{DeU-Net}
\author{Shunjie Dong\inst{1} \and  
		Jinlong Zhao\inst{1} \and  
		Maojun Zhang\inst{1} \and  
		Zhengxue Shi\inst{1} \and  
		Jianing Deng\inst{1} \and  
		Yiyu Shi\inst{2} \and      
		Mei Tian\inst{1} \and      
		Cheng Zhuo\inst{1}\textsuperscript{(\Letter)}}  
%
\authorrunning{S. Dong et al.}
%
\institute{Zhejiang University, Hangzhou, China\\
\email{\{sj\_dong,zhaojl,zhmj,sjwo,dengjn,meitian,czhuo\}@zju.edu.cn}\and
University of Notre Dame, Notre Dame, USA\\
\email{yshi4@nd.edu}}

\maketitle              
\begin{abstract}
Automatic segmentation of cardiac magnetic resonance imaging (MRI) facilitates efficient and accurate volume measurement in clinical applications. However, due to anisotropic resolution and ambiguous border (e.g., right ventricular endocardium), existing methods suffer from the degradation of accuracy and robustness in 3D cardiac MRI video segmentation. In this paper, we propose a novel \textit{Deformable U-Net} (DeU-Net) to fully exploit spatio-temporal information from 3D cardiac MRI video, including a Temporal Deformable Aggregation Module (TDAM) and a Deformable Global Position Attention (DGPA) network. First, the TDAM takes a cardiac MRI video clip as input with temporal information extracted by an offset prediction network. Then we fuse extracted temporal information via a temporal aggregation deformable convolution to produce fused feature maps. Furthermore, to aggregate meaningful features, we devise the DGPA network by employing deformable attention U-Net, which can encode a wider range of multi-dimensional contextual information into global and local features. Experimental results show that our DeU-Net achieves the state-of-the-art performance on commonly used evaluation metrics, especially for cardiac marginal information (ASSD and HD). 
\end{abstract}

\section{Introduction}
Magnetic Resonance Imaging (MRI) is widely used by cardiologists as the golden modality for cardiac assessment~\cite{vick2009gold}. The segmentation of kinetic MR images along the short axis is complicated but essential to precise morphological and pathological analysis, diagnosis, and surgical planning. In particular, one has to delineate left ventricular endocardium (LV), myocardium (MYO), and right ventricular endocardium (RV) to calculate the volume of the cavities in cardiac MRI video, including end-diastolic (ED) and end-systolic (ES) phases~\cite{peng2016review}.

Recent studies~\cite{ronneberger2015u,oktay2018attention,zheng2019hfa,zotti2017gridnet,wang2019msu,Deng2020STDF} proposed deep learning based approaches to learn robust contextual and semantic features, achieving the state-of-the-art segmentation performance. However, automatic and accurate 3D cardiac MRI video segmentation still remains very challenging due to significant variations in the subjects, ambiguous borders, inhomogeneous intensity and artifects, especially for RV. There are two key issues that needs to be resolved: (1) In cardiac MRI video, the RV segmentation performance is influenced by complicated shape and inhomogeneous intensity, especially the partial volume effect close to the free wall. (2) Subtle structures (e.g., MYO) have ambiguous borders and different orientations in different anatomical planes of MRI video, causing segmentation inaccuracy. Thus, it is highly desired to have precise and robust cardiac MRI video segmentation.

In this paper, we propose a new \textit{Deformable U-Net} (DeU-Net) to address the aforementioned issues by fully exploiting the spatio-temporal information from 3D cardiac MRI video and aggregating temporal information to boost segmentation performance. The DeU-Net consists of two parts: Temporal Deformable Aggregation Module (TDAM) and a Deformable Global Position Attention (DGPA) network. To address the partial volume effect of RV in~\cite{zheng2019hfa,zotti2017gridnet}, the TDAM utilizes the spatio-temporal information of the MRI video clip to produce fused feature maps by a temporal aggregation deformable convolution. To handle the issue of subtle structures in~\cite{oktay2018attention}, the U-Net based DGPA network jointly encodes a wider range of multi-dimensional contextual information into global and local features, guaranteeing clear and continuous borders of every segmentation map. The experimental results quantitatively and qualitatively show that our proposal achieves the state-of-the-art performance on commonly used metrics, especially for cardiac marginal information (ASSD and HD). 


\section{Method}
\begin{figure}[t]
\centering
\includegraphics[width=1\textwidth]{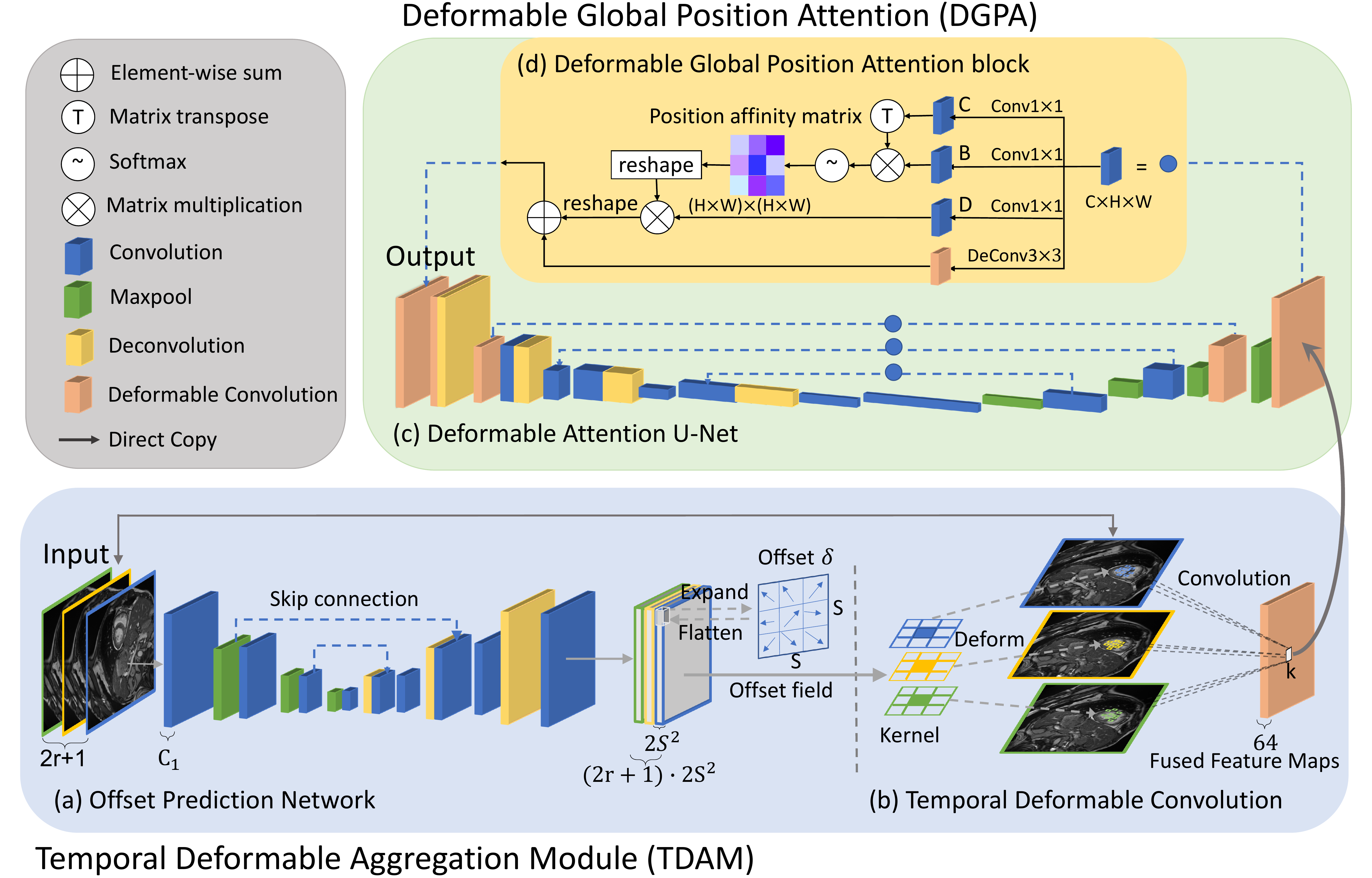}
\caption{The architecture of DeU-Net for 3D cardiac MRI video segmentation. Given a video clip ($2r+1$ concatenated frames) as input, an offset prediction network is designed for deformable offset. Temporal deformable aggregation convolution exploits the offset field to fuse temporal information. The fused feature maps are used by a deformable attention U-Net to enhance segmentation performance. Herein, temporal radius $r=1$ and deformable kernel size $S=3$.} 
\label{network}
\end{figure}

The architecture of DeU-Net is plotted in Fig.~\ref{network}, including a Temporal Deformable Aggregation Module (TDAM) and a Deformable Global Position Attention (DGPA) network. The proposed TDAM consists of two phases: a temporal deformable convolution and an offset prediction network based on U-Net to predict deformable offsets. The fused features produced by TDAM are fed into DGPA for the final segmentation results. The DGPA network which also employs U-Net as backbone introduces deformable convolution for encoders and utilizes the deformable attention block to augment the spatial sampling locations.

\subsection{Temporal Deformable Aggregation Module}

Many existing methods design very complicated neural networks to achieve performance gain. However, most approaches ignore the spatio-temporal information of 3D MRI video, and treat each frame as a separate object, thereby causing performance degradation. Moreover, in the process of data sampling, various semantic details of the video clips may get lost due to fast variation of cardiac borders and regular convolution, inevitably distorting video local details and pixel-wise connections between the frames. Thus, we propose a Temporal Deformable Aggregation Module (TDAM) to adaptively extract temporal information (motion field) for image interpretation. The proposed TDAM takes a target frame along with its neighboring reference frames as inputs to jointly predict an offset field. Then, the enhanced contextual information can be fused into the target frame by a temporal aggregation deformable convolution.
\begin{figure}[t]
\centering
\includegraphics[width=1\textwidth]{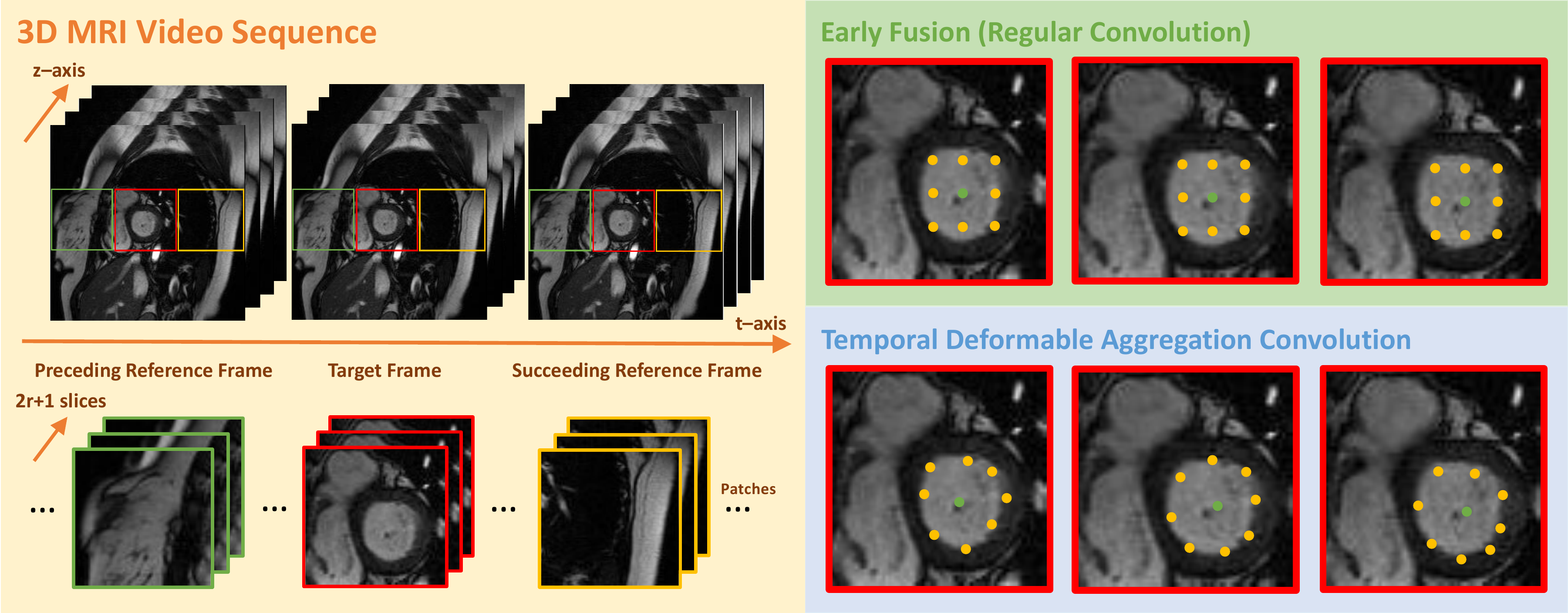}
\caption{Comparison between the fixed receptive field in Early Fusion~\cite{karpathy2014large} and the adaptive receptive field in TDAM. Each canonical clip is extracted using the samples from the same position in the $x-y$ plane. The yellow points denote the sampling positions of $3\times3$ convolution window centered at green points. The temporal change of cardiac borders are highlighted, corresponding to the relevant context in the 3D MRI video. The patches are the images collected at different timestamps with the same position.} 
\label{deform_conv}
\end{figure}

We denote the dimension of an input 3D MRI video as $[x, y, z, t]$, where $x-y$ plane is the short-axis plane, $z$ is the short axis, and $t$ is the temporal dimension. Considering the large inter-slice gap in MRI cardiac images along $z$ axis, we select images within the dimension $[x, y, t]$ where stronger correlation may exist. Specially, we define $C_{t_0} \in \mathbb{R}^{H \times W}$ as the target frame in a 3D MRI video at time $t_0$, where $H$ and $W$ are the height and width of the input feature map. In order to leverage temporal information, we take the preceding and succeeding $r$ frames as the reference to improve the quality of the target frame $C_{t_0}$. For a 3D cardiac MRI video clip $\left\{ C_{t_0-r},\cdots,C_{t_0},\cdots,C_{t_0+r}\right\}$, the conventional temporal fusion scheme (i.e., Early Fusion~\cite{karpathy2014large}) can be formulated as a multichannel convolution directly applied on the target frames as:\vspace{-0.2cm}
\begin{equation}
\label{TDAM}
  \mathcal{F}(\mathbf{k}) = \sum_{t={t_0-r}}^{t_0+r}\sum_{s=1}^{S^2}K_{t,s}\cdot C_t(\mathbf{k}+\mathbf{k}_s),\vspace{-0.2cm}
\end{equation}
where $\mathcal{F}(\cdot)$ represents the quality-enhanced feature map. $S$ is the size of convolution kernel and $K_t \in \mathbb{R}^{S^2}$ denotes the kernel for $t$-th channel. $\mathbf{k}$ represents an arbitrary spatial position and $\mathbf{k}_s$ indicates the regular sampling offsets. However, noisy content can be easily introduced due to cardiac temporal change in video. Inspired by Multi-Frame Quality Enhancement~\cite{yang2018multi} for Video Quality Enhancement (VQE), we design TDAM to augment the regular sampling offset with extra learnable offset $\mathbf{\delta}_{(t,\mathbf{k})} \in \mathbb{R}^{2S^2}$ for the potential spatio-temporal correlations:\vspace{-0.1cm}
\begin{equation}
  \mathbf{k}_s \leftarrow \mathbf{k}_s + \mathbf{\delta}_{(t,\mathbf{k}),s}.\vspace{-0.1cm}
\end{equation}
Note that the deformable offset $\mathbf{\delta}_{(t,\mathbf{k})}$ is designed for each convolution window centered at a spatio-temporal position $(t,\mathbf{k})$, as shown in Fig.~\ref{network}. We propose to take the whole clip into consideration and jointly predict all the deformable offsets with a U-Net based network~\cite{ronneberger2015u}. Maxpool and deconvolutional layers are used for downsampling and upsampling, respectively. Convolutional layer with stride $1$, zero paddings are designed to retain the feature size. With such a scheme, the spatial deformations with temporal dynamics in 3D MRI video clips can be simultaneously modeled.

Compared with the existing VQE approaches which achieve explicit motion compensation before fusion to alleviate the negative effects of temporal motion, TDAM implicitly focuses on cardiac border cues with position-specific sampling. As shown in Fig.~\ref{deform_conv}, adjacent deformable convolution windows can independently sample the contents, achieving higher flexibility and performance boost.

\subsection{Deformable Global Position Attention.}
Regular convolution is restricted by the kernel size as well as the fixed geometric structures, which results in limited performance in modeling geometric transformations. In practice, it is very difficult to reduce false-positive predictions due to unclear borders between the cardiac instances. To address these issues, we propose a Deformable Global Position Attention (DGPA) network to capture a sufficiently large receptive field and semantic global contextual information. DGPA augments the spatial sampling locations in the modules with additional offsets, which is designed to model complex geometric transformations. Thus, long-range contextual information can be collected, which helps obtain more discriminative cardiac border for pixel-level prediction.

As shown in Fig.~\ref{network}, the fused local features $I \in \mathbb{R}^{N\times H\times W}$ are regarded as input to the DGPA block, where $N$ represents the number of input channels, $H$ and $W$ indicate the height and width of the input features, respectively. We first feed the input features $I$ with a $3\times 3$ deformable convolution layer to capture cardiac geometric information. The formulation is as below:\vspace{-0.15cm}
\begin{equation}
  O = I \otimes \mathcal{K}_{l,\delta'}, \vspace{-0.15cm}
\end{equation}
where $O\in\mathbb{R}^{N\times H\times W}$ the feature map, $\mathcal{K}$ is deformable convolution kernel, $l$ is the kernel size and $\delta'$ is the deformable offset. The input feature map is reshaped to three new feature maps ${B,C,D}\in \mathbb{R}^{N\times M}$, where $M$ denotes the number of pixels ($M=H\times W$). In order to exploit the high-level features of cardiac borders, a dot-product is conducted between $B$ and the transpose of $C$. Then the result is applied into a softmax layer to calculate the attention map $P\in\mathbb{R}^{N\times N}$:\vspace{-0.15cm}
\begin{equation}
  p_{ji} = \frac{\exp(B_{i} \cdot C^{T}_{j})}{\sum^{N}_{i=1} \exp(B_{i} \cdot C^{T}_{j})},\vspace{-0.15cm}
\end{equation}
where $p_{ji}$ represents the $i^{th}$ pixel's impact on the $j^{th}$ pixel. The more similar feature representations of the two pixels indicate a stronger correlation between them. Then we perform a matrix multiplication between the transpose of $P$ and $D$ to reshape the result to $\mathbb{R}^{N\times H\times W}$. Finally, an element-wise sum operation is applied with the feature map $O$ from deformable block to obtain the output features $\mathcal{Z}\in \mathbb{R}^{N\times H\times W}$ as below:\vspace{-0.2cm}
\begin{equation}
  \mathcal{Z}_{j} = \alpha \sum^{N}_{i=1} (p_{ji}\cdot D_{i})+O_{j},\vspace{-0.2cm}
\end{equation}
where $\alpha$ is the scale parameter belonging to the position affinity matrix. Each element in $\mathcal{Z}$ is a weighted sum of the features globally and selectively aggregates input features $I$. Long-range dependencies of the feature map are calculated to improve intra-class compact and semantic consistency. 


\section{Experiments and Results}

\subsection{Setup}

\paragraph{Evaluation Datasets.} We evaluate our proposal and competitive approaches on the publicly available data of ACDC MICCAI 2017 Challenge~\cite{bernard2018deep} with additional labeling done by experience radiologists~\cite{wang2019msu}. The dataset has right ventricle, myocardium, and left ventricle segmentation frames from MRI videos with labels provided by experienced radiologists. The collected images following the common clinical SSFP cine-MRI sequence have similar properties as 3D cardiac MRI videos. The MRI sequence, as a series of short-axis slices of end diastolic and end systolic instant, starts from the mitral valves down to the apex of the left ventricle. We resize the exams into 256 $\times$ 256 images, and no additional pre-processing was conducted. The dataset has 150 exams from different patients with 100 for training and 50 for testing.\vspace{-6pt}

\paragraph{Implementation Details.} The proposed method is implemented based on PyTorch library with reference to MMDetection toolbox~\cite{chen2019mmdetection} for deformable convolution, using a NVIDIA GTX 1080Ti GPU. For the training set, standard data augmentation (i.e., mirror, axial flip or rotation) is further used to exploit training samples better. We use Adam optimizer to update the network parameters. The initial learning rate is set to $2\times 10^{-4}$ and a weight decay of $1\times10^{-4}$. We use a batch size of at least $12$. The number of reference frames $r$ from Eq.~\ref{TDAM} is set as $1$. The training is stopped if the Dice score does not increase by 20 epochs. In our experiments, we perform 5-fold cross-validation.\vspace{-6pt}

\paragraph{Comparison Methods.} We compare the proposed DeU-Net with several state-of-the-art approaches for 3D Cardiac MRI Video segmentation: (1) \textit{k}U-Net~\cite{chen2016combining} combined convolutional and recurrent neural networks to exploit the intra-slice and inter-slice contexts, respectively. (2) GridNet~\cite{zotti2017gridnet} incorporated a shape prior whose registration on the input image is learned by the model, learning both high-level and low-level features. (3) Attention U-Net~\cite{oktay2018attention} proposed a novel self-attention gating module to learn irrelevant regions in an input image for dense label predictions. For a fair comparison, all these methods are modified and fully trained for 3D cardiac MRI video segmentation.\vspace{-6pt}

\subsection{Results} 

For quantitative evaluation, Table~\ref{Dice} details the comparison among U-Net, GridNet, Attention U-Net, \textit{k}U-Net, and the proposed DeU-Net on average symmetric surface distance (ASSD), Hausdorff distance (HD), and Dice score. In addition to the scores at ED and ES phases, we evaluate the average score of the entire 3D cardiac MRI video for all the three metrics (with more detailed statistics reported in the supplementary materials due to the space limit). We can observe that DeU-Net significantly outperforms all the prior networks on most metrics. It is worth noting that our proposal substantially improves segmentation performance on RV, where the ventricle has complicated shape and intensity inhomogeneities. Note that, the proposed method achieves the best results on ASSD that is lower than GridNet by an average of $0.19$ mm. Compared to existing methods, HD is lower for our approach by an average of $7.15$ mm, and the Dice score is better by an average of $5\%$. This is a strong indication that DeU-Net exploits spatio-temporal information and reduces the negative effect of cardiac border ambiguity.

To separate the contributions of TDAM and DGPA, we also evaluate the performance of three variants of DeU-Net: (1) DeU-Net(t), the variant without TDAM. (2) DeU-Net(d), the variant without DGPA. (3) ToFlow U-Net, the variant replacing TDAM by Task-oriented Flow (ToFlow)~\cite{xue2019video}. For the differences among RV, MYO and LV, Table~\ref{Dice} shows that DeU-Net(d) works better than ToFlow U-Net and DeU-Net(t) on both HD and ASSD, indicating that TDAM can fully explore temporal correspondences across multiple frames, especially close to the borders. Moreover, DeU-Net also performs better over all the three variants, validating the necessity of having both TDAM and DGPA in the flow.
\begin{table}[tbp]
  \centering
  \caption{Average scores of the 3D cardiac MRI video different metrics and approaches.}
  \label{Dice}
  \resizebox{\textwidth}{!}{
    \begin{tabular}{|l|c|c|c|c|c|c|c|}
    \hline
    \multirow{2}{*}{Method}&\multicolumn{2}{c}{ASSD LV}&\multicolumn{2}{c}{ASSD MYO}&\multicolumn{2}{c}{ASSD RV}&ASSD\\
    \cline{2-8}
    &\multicolumn{1}{c}{ED}&ES&\multicolumn{1}{c}{ED}&ES&\multicolumn{1}{c}{ED}&ES&Average\\
    \hline
    U-Net                    &0.34$\pm$.09      &0.51$\pm$.08      &0.36$\pm$.07       &0.41$\pm$.05     &0.81$\pm$.06      &1.65$\pm$.07     &0.78$\pm$.06\\    
    \textit{k}U-Net          &0.20$\pm$.12      &0.32$\pm$.07      &0.34$\pm$.03       &0.41$\pm$.07     &0.62$\pm$.08      &0.58$\pm$.06     &0.51$\pm$.11\\
    GridNet                  &0.16$\pm$.10      &0.25$\pm$.09      &0.24$\pm$.07       &0.26$\pm$.09     &0.27$\pm$.11      &0.49$\pm$.08     &0.38$\pm$.08\\
    Attention U-Net          &0.17$\pm$.13      &0.32$\pm$.11      &0.24$\pm$.09       &0.27$\pm$.09     &0.26$\pm$.08      &0.56$\pm$.13     &0.39$\pm$.12\\
    \hline
    ToFlow U-Net              &0.12$\pm$.07      &0.28$\pm$.04      &0.21$\pm$.05       &0.15$\pm$.04     &0.30$\pm$.06      &0.57$\pm$.08     &0.27$\pm$.10\\
    DeU-Net(t)               &0.07$\pm$.08      &0.23$\pm$.09      &0.19$\pm$.02       &0.18$\pm$.06     &0.22$\pm$.04      &0.51$\pm$.09     &0.24$\pm$.08\\
    DeU-Net(d)               &0.10$\pm$.04      &0.18$\pm$.06      &0.17$\pm$.03       &0.20$\pm$.03     &0.19$\pm$.05      &0.49$\pm$.02     &0.22$\pm$.05\\
    DeU-Net                  &{\bf 0.04$\pm$.01}&{\bf 0.12$\pm$.04}&{\bf 0.12$\pm$.02}&{\bf 0.12$\pm$.00}&{\bf0.13$\pm$.03}&{\bf 0.41$\pm$.03}&{\bf 0.19$\pm$.04}\\
    \hline
    \hline
    \multirow{2}{*}{Method}&\multicolumn{2}{c}{HD LV}&\multicolumn{2}{c}{HD MYO}&\multicolumn{2}{c}{HD RV}&HD\\
    \cline{2-8}
    &\multicolumn{1}{c}{ED}&ES&\multicolumn{1}{c}{ED}&ES&\multicolumn{1}{c}{ED}&ES&Average\\
    \hline
    U-Net                    &6.17$\pm$.88     &8.29$\pm$.62      &15.26$\pm$.98    &17.92$\pm$.23     &20.51$\pm$.54    &21.21$\pm$.88       &19.89$\pm$.73\\
    \textit{k}U-Net          &4.59$\pm$.32     &5.40$\pm$.55      &7.11$\pm$.79     &6.45$\pm$.64      &11.92$\pm$.52    &14.83$\pm$.42       &14.38$\pm$.70\\
    GridNet                  &5.96$\pm$.42     &6.57$\pm$.41      &8.68$\pm$.45     &8.99$\pm$.84      &13.48$\pm$.53    &16.66$\pm$.83       &15.06$\pm$.25\\
    Attention U-Net          &4.39$\pm$.76     &5.27$\pm$.74      &7.02$\pm$.88     &7.35$\pm$.90      &12.65$\pm$.58    &10.99$\pm$.41       &13.95$\pm$.21\\
    \hline
    ToFlow U-Net              &3.21$\pm$.47      &4.32$\pm$.63      &6.20$\pm$.91       &6.02$\pm$.69      &10.39$\pm$.81      &13.00$\pm$.24    &11.19$\pm$.43\\
    DeU-Net(t)               &4.08$\pm$.32      &4.59$\pm$.09      &5.91$\pm$.33       &5.31$\pm$.70      &11.58$\pm$.12      &12.23$\pm$.13    &9.28$\pm$.14\\
    DeU-Net(d)               &2.76$\pm$.36      &4.27$\pm$.55      &4.77$\pm$.40       &4.22$\pm$.43      &12.13$\pm$.23      &11.97$\pm$.09    &7.69$\pm$.24\\
    DeU-Net                  &{\bf 2.48$\pm$.27}&{\bf 3.25$\pm$.30}&{\bf 4.56$\pm$.29} &{\bf4.20$\pm$.14} &{\bf 9.88$\pm$.17} &   {\bf9.02$\pm$.11}    &{\bf 6.80$\pm$.17}\\
    \hline
    \hline
    \multirow{2}{*}{Method}&\multicolumn{2}{c}{Dice LV}&\multicolumn{2}{c}{Dice MYO}&\multicolumn{2}{c}{Dice RV}&Dice\\
    \cline{2-8}
    &\multicolumn{1}{c}{ED}&ES&\multicolumn{1}{c}{ED}&ES&\multicolumn{1}{c}{ED}&ES&Average\\
    \hline
    U-Net                    &0.96$\pm$.00      &0.90$\pm$.01      &0.78$\pm$.01    &0.76$\pm$.02    &0.88$\pm$.02      &0.80$\pm$.02      &0.81$\pm$.03\\
    \textit{k}U-Net          &0.96$\pm$.00      &0.90$\pm$.00      &0.88$\pm$.02    &0.89$\pm$.03    &0.91$\pm$.03      &0.82$\pm$.03      &0.83$\pm$.01\\
    GridNet                  &0.96$\pm$.01      &0.91$\pm$.01      &0.88$\pm$.03    &0.90$\pm$.03    &0.90$\pm$.01      &0.82$\pm$.03      &0.85$\pm$.02\\
    Attention U-Net          &0.96$\pm$.01      &0.91$\pm$.02      &0.88$\pm$.02    &0.90$\pm$.01    &0.91$\pm$.01      &0.83$\pm$.02      &0.84$\pm$.03\\
    \hline
    ToFlow U-Net              &0.96$\pm$.00      &0.91$\pm$.01      &{\bf 0.90$\pm$.01}&0.90$\pm$.01      &0.92$\pm$.01      &0.84$\pm$.02      &0.87$\pm$.02\\
    DeU-Net(t)               &0.96$\pm$.00      &0.91$\pm$.00      &0.89$\pm$.00      &0.90$\pm$.00      &0.92$\pm$.01      &0.83$\pm$.01      &0.86$\pm$.03\\
    DeU-Net(d)               &0.96$\pm$.01      &0.91$\pm$.01      &0.88$\pm$.01      &{\bf 0.91$\pm$.01}&0.92$\pm$.00      &0.84$\pm$.00      &0.88$\pm$.02\\
    DeU-Net                  &{\bf 0.97$\pm$.00}&{\bf 0.92$\pm$.00}&{\bf 0.90$\pm$.00}&{\bf 0.91$\pm$.01}&{\bf 0.93$\pm$.00}  &{\bf 0.86$\pm$.01}      &{\bf 0.90$\pm$.01}\\
    \hline
    \end{tabular}
    }
    \vspace{-0.5cm}
\end{table}

\begin{figure}[!tbp]
\centering
\includegraphics[width=0.9\textwidth]{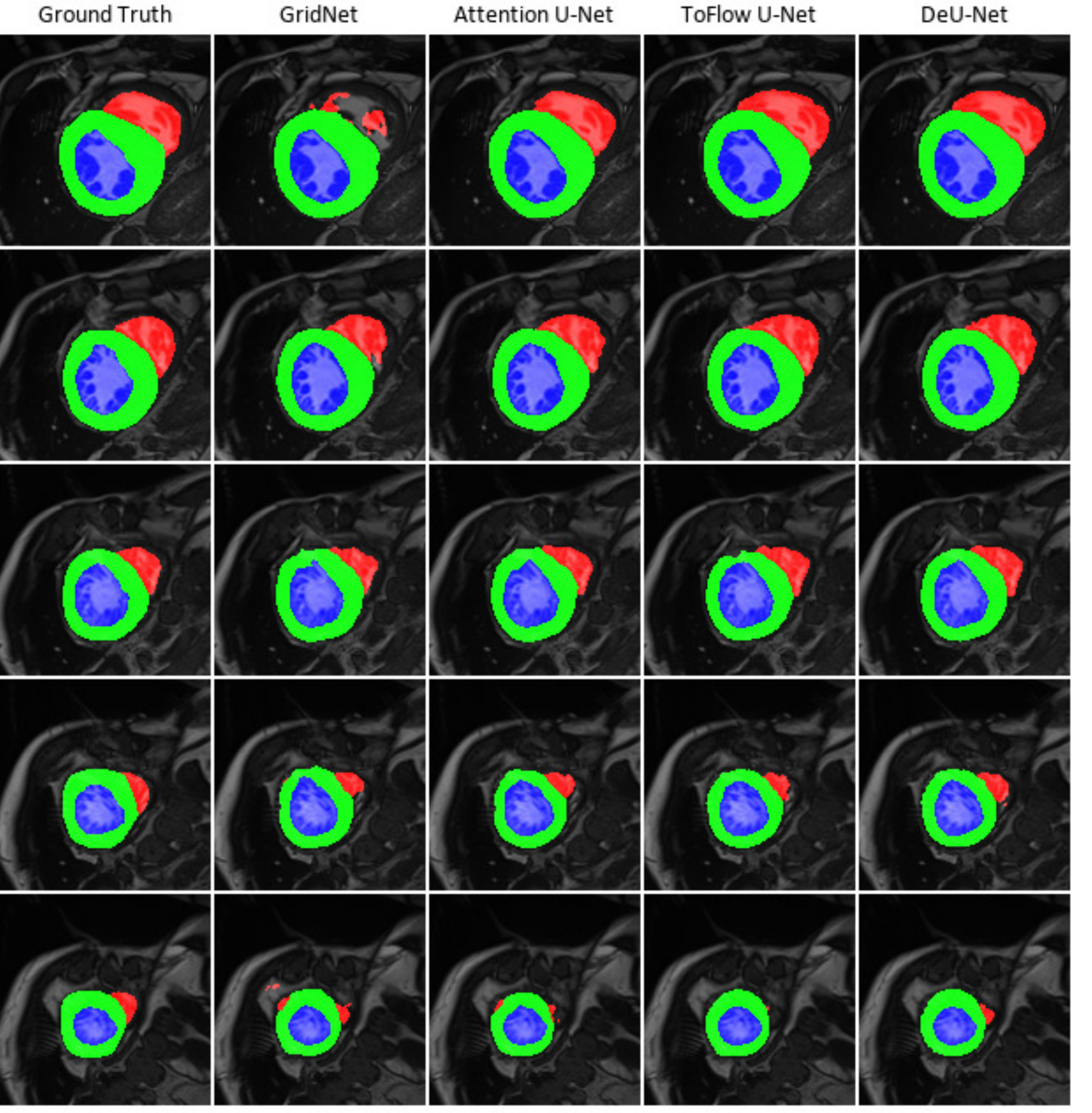}
\vspace{-10pt}
\caption{Visualization of the segmentation results by different methods on the testing data. RV, MYO, and LV are labeled in red, green and blue, respectively.} 
\label{segmentation}
\vspace{-8pt}
\end{figure}

Finally, Fig.~\ref{segmentation} illustrates a visual comparison of the groud truth, GridNet, Attention U-Net, ToFlow U-Net and DeU-Net in a 3D cardiac MRI video clip. It can be seen that GridNet and Attention U-Net successfully produce accurate results on most slices of each 3D volume, but the shape of the target region is not as accurate as ToFlow U-Net and DeU-Net. Such observations are especially apparent in the rows of 2, 3, and 4 of Fig.~\ref{segmentation}. Moreover, DeU-Net accurately extracts the borders of the target regions on most of the slices, as shown in the last row of Fig.~\ref{segmentation}. Note that, the segmentation performance of RV, labeled as blue, is significantly lower than that of MYO and LV due to the irregular shape and the ambiguous borders.

\section{Discussions and Conclusions}
In this paper, we propose a \textit{Deformable U-Net} (DeU-Net) to fully exploit spatio-temporal information from 3D cardiac MRI video, including a Temporal Deformable Aggregation Module (TDAM) and a Deformable Global Position Attention (DGPA) network. Based on the temporal correlation across multiple frames, TDAM aggregates temporal information with learnable sampling offsets, and capture sufficient semantic context. To obtain the discriminative and compact features in subtle structures, the DGPA network encodes a wider range of multi-dimensional fused contextual information into global and local features. Experimental results show that our proposal achieves the state-of-the-art performance on commonly used metrics, especially for cardiac marginal information (ASSD and HD). In the future, it would be of interest to apply our proposal to other datasets (such as myocardial contrast echocardiography). Our segmentation method will facilitate the translation of neural networks to clinical practice.

\subsubsection{Acknowledgement.} This work was supported in part by National Key Research and Development Program Program of China [No. 2018YFE0126300], Key Area Research and Development Program of Guangdong Province [No. 2018B030338001], and Information Technology Center, Zhejiang University.
\bibliographystyle{splncs04}
\bibliography{paper910}

\end{document}